# $^2$H-NMR studies of supercooled and glassy aspirin

R. Nath, B. Geil, R. Böhmer

*Fachbereich Physik and Interdisziplinäres Zentrum für Magnetische Resonanz,*

*Universität Dortmund, 44221 Dortmund, Germany*

Acetyl salicylic acid, deuterated at the methyl group, was investigated using $^2$H-NMR in its supercooled and glassy states. Just above the glass transition temperature the molecular reorientations, studied using stimulated-echo spectroscopy, demonstrated a large degree of similarity with other glass formers. Deep in the glassy phase the NMR spectra look similar to those reported for the crystal [A. Detken, P. Focke, H. Zimmermann, U. Haeberlen, Z. Olejniczak, Z. T. Lalowicz, Z. Naturforsch. A 50 (1995) 95] and below 20 K they are indicative for rotational tunneling with a relatively large tunneling frequency. Measurements of the spin-lattice relaxation times for temperatures below 150 K reveal a broad distribution of correlation times in the glass. The dominant energy barrier characterizing the slow-down of the methyl group is significantly smaller than the well defined barrier in the crystal.

# 1. Introduction

There are numerous examples for relaxations in complex systems, with the molecular dynamics in viscous and vitreous pharmaceuticals being a relatively recent issue that attracts interest in the scientific community dealing with supercooled liquids and glasses. In view of the polymorphism and hence potential variability of properties that many crystalline pharmaceuticals exhibit [1] the amorphous forms of these substances are important to study. Further motivation is provided by the sometimes increased physiological dissolubility of the vitreous phase and the possibility of pharmaceutical alloying [2].

Amorphous aspirin has its glass transition temperature, $T_g$, at 243 K, i.e., much below ambient temperature, and is therefore of no practical importance. But the supercooled liquid and glassy states of aspirin were nevertheless investigated quite frequently, e.g., using calorimetry [3,4,5], dielectric measurements [6,7], light scattering [8], and deuteron NMR [7]. Our previous NMR study on methyl deuterated samples essentially focused on measurements of the spin-lattice relaxation for $T > 70$ K [7], also employing specially designed pulse sequences [9].

In the crystalline modification the rotational tunneling of the $CD_3$ group, observable at temperatures below about 30 K, is a major concern and has been investigated via $^{13}$C-NMR [10] and particularly thoroughly via $^2$H-NMR [11,12,13]. The methyl group tunneling was not only studied in a variety of crystals [14] but also in glasses, see e.g., [15]. While it is clear that the tunneling motion can be suppressed by *intra*molecular potentials as, e.g., in glassy propylene carbonate [16] or in molecules with partially deuterated methyl groups [17], it has been argued that the *inter*molecular potential should not play a role in this respect.

One goal of the present article is to provide a detailed comparison of the low-temperature results for glassy aspirin with those obtained for the crystalline analog [11,12]. We find that the deuteron spectra for both forms are very similar, while the spin-lattice relaxation times differ considerably. In addition to these studies of the methyl-group dynamics, we carried out stimulated-echo measurements near $T_g$ in order to clarify the geometry of the overall molecular reorientation in this glass former. This complements our previous study on the slow dynamics of supercooled liquid aspirin [7].

This paper is organized as follows. First, we provide some experimental details. Then, in Sections 3.1 and 3.2 we present deuteron spectra and spin-relaxation times as obtained deeply in the glassy phase. The reorientational motion of the aspirin molecule just above the glass transition is dealt with in Section 3.3. Finally, in Section 4 we discuss our results and in Section 5 summarize the findings.

## 2. Experimental details

The methyl deuterated acetyl salicylic acid (aspirin-$d_3$) used for the present work is the same as studied previously [7]. The crystal powder was sealed in an evacuated NMR tube which was then slowly heated above the melting point (~ 408 K). The molten sample was subsequently quenched in ice water to suppress crystallization. The NMR measurements were carried out in a continuous flow cryostat using a home-built spectrometer which was operated at a Larmor frequency of 46.46 MHz. The length of the $\pi/2$-pulse typically was 3.5 μs. Longitudinal magnetization decay curves were measured using saturation recovery techniques augmented by a solid echo. Also the transversal magnetization decay was monitored using a solid-echo pulse sequence. The reported solid-echo spectra were measured with a pulse echo delay of $t_{SE}$ = 50 μs. Furthermore, stimulated-echo decay curves were recorded with four-pulse sequences at several temperatures somewhat above $T_g$.

## 3. Results

### 3.1 Solid-echo spectra

Solid-echo spectra of vitreous aspirin-$d_3$ were obtained at several temperatures and are shown in Fig. 1. At high temperatures the spectra are practically indistinguishable from those reported for the crystal [18]. These spectra can be described by the powder average of the quadrupolar frequency $\omega_Q$. For each CD bond with a well defined orientation the quadrupolar splitting $\omega_Q$ is given by

$$\omega_Q = \tfrac{3}{8}(e^2qQ/\hbar)(3\cos^2\theta - 1 - \eta\sin^2\theta\cos2\phi). \tag{1}$$

Here the angles $\theta$ and $\phi$ describe the orientation of the EFG tensor with respect to that of the external static magnetic field in the usual fashion. In the case of an axially symmetric electric field gradient (EFG) tensor with $\eta = 0$ Eq. (1) reduces to $\omega_Q = \delta \times P_2(\cos\theta)$ with $P_2(\cos\theta) = \tfrac{1}{2}(3\cos^2\theta - 1)$. $\delta$ is called the anisotropy parameter and $\eta$ the asymmetry parameter.

The parameters determined for the methyl group in the crystal are $\delta_M = 2\pi \times 43.5$ kHz and $\eta_M = 0.1$ [18]. They also apply to the spectra of glassy aspirin at temperatures $T > 30$ K, see Fig. 1. The numerical value of the anisotropy parameter is typical for a quickly rotating methyl group, while the nonzero asymmetry parameter is somewhat unusual. On the basis of measurements of partially relaxed solid-echo spectra librational motions of the threefold axis of the $CD_3$ group could be ruled out as a possible explanation [18]. Merely, electronic effects of the oxygen atom adjacent to the methyl group were held responsible for the nonzero asymmetry parameter [18].

Over a wide temperature range the shape of the spectra remains the same, only for $T \leq 30$ K it changes significantly. As highlighted by the dashed line in Fig. 1, at ±63.5 kHz a spectral feature emerges which becomes more pronounced at lower temperatures. At first glance, the spectra look like so called two-phase spectra [19]. The broad component would then correspond to $CD_3$ groups, the dynamics of which is frozen on the time scale of about $1/\delta_M \approx 3.7$ μs.

In fact, the broad component of the spectra can be described by taking the powder average of Eq. (1) with $\delta_Q = 2\pi \times 131$ kHz and $\eta_Q = 0$. A superposition of the corresponding Pake pattern with a calculated methyl spectrum [20] using $\delta_M = 2\pi \times 43.5$ kHz and $\eta_M = 0.1$ is shown in Fig. 2 as dashed line. At 12 K the weight of the broad component is about one third of that of the integrated spectral intensity. Within this two-phase interpretation the opening angle $\gamma$ of the $CD_3$ group can be obtained from $P_2(\cos\gamma) = -\delta_M/\delta_Q$. From the shape of the experimental spectra, which is essentially reproduced by the dashed line shown in Fig. 2, we find $\gamma \approx 70.1°$. This value is quite close to (the complement of) the tetrahedral angle of $\gamma_{tet}$ = 70.53° corresponding to $P_2(\cos\gamma_{tet}) = -1/3$.

The discussion provided so far ignores that methyl groups can be subject to rotational tunneling at low temperatures. As we will argue below the relevant tunneling frequency $\omega_t$ is much larger than $\omega_Q$. In this case a set of five doublets is expected for a general orientation of the threefold symmetry axis of the methyl group. These 10 lines appear at frequencies $\pm\beta$, $\pm(|\alpha| \pm \beta)$, and $\pm(2|\alpha| \pm \beta)$ with [11,21,22]

$$\beta_{tet} = \tfrac{1}{8}(e^2qQ/\hbar)(3\cos^2\theta - 1) \tag{2}$$

and

$$|\alpha_{tet}| = \tfrac{1}{4}(e^2qQ/\hbar)\sin\theta\sqrt{1+\cos^2\theta - \sqrt{2}\sin 2\theta \cos 3\phi}. \tag{3}$$

Here the subscript 'tet' is again meant to indicate that these equations refer to the assumption that the $CD_3$ group is characterized by the tetrahedral angle and thus $\beta_{tet} = \omega_Q/3$. The expressions for general $\gamma$ are given in [11]. In Eq. (3) the angle $\phi$ refers to the azimuthal angle of the external magnetic field in the plane perpendicular to the $C_3$ axis of the methyl group, for a sketch of the molecular coordinate frame see Fig. 2 of [11]. The form $\cos(3\phi)$ in the expression for $\alpha_{tet}$ reflects the occurrence of a threefold potential, the nature of which is discussed below. This form also implies that it is irrelevant which of the three CD bonds is used to define the x-axis in the molecular frame associated with the methyl group.

The five doublets at the frequencies $\pm\beta$, $\pm(|\alpha| \pm \beta)$, and $\pm(2|\alpha| \pm \beta)$ are characterized by relative intensities of 12:2:2:1:1 [12]. For powdered or amorphous specimens this means that the inner "methyl" pattern, calculated using Eq. (2), should theoretically account for 2/3 of the total intensity. It has to be pointed out, however, that the powder average computed using Eq. (2) and (3) can only describe simple spectra obtained by the Fourier transform of the free induction decay (FID). These are formally obtained from solid-echo spectra for a vanishing echo delay, $t_{SE} \to 0$.

Spectra for finite $t_{SE}$ were computed *numerically* for powders of aspirin-$d_3$ [12] employing $e^2qQ/\hbar$ = 170 kHz and $\omega_t = 2\pi \times 2.59$ MHz. Some of these spectra are reproduced in Fig. 2. In particular the spectrum calculated for $t_{SE}$ = 42 µs agrees reasonably well with our glass data. This suggests that the tunneling of the methyl group is insensitive to whether or not it is embedded into a crystalline or into an amorphous environment.

In the crystal the tunneling frequency is temperature dependent and for T < 25 K it was found to be $\omega_t > 2\pi \times 1.5$ MHz $\gg \omega_Q$ [11]. Therefore, the more complicated case of intermediate tunneling frequencies, $\omega_t \sim \omega_Q$, associated with a "forest of lines" [11] is not discussed here. For $\omega_t \ll \omega_Q$ the classical limit corresponding to a rigid rotor is reached and the calculated spectrum is composed of 6 lines [11], i.e., one pair of lines for each of the three CD bonds in a methyl group.

## 3.2 Spin-spin and spin-lattice relaxation

The longitudinal magnetization M(t) for amorphous aspirin-$d_3$ as measured subsequent to saturation is presented in Fig. 3 for several temperatures. The M(t) curves are normalized such that they decay from 1 to 0. For T > 30 K the decay time, i.e., the spin-lattice relaxation time $T_1$, decreases with decreasing temperatures, and for T < 30 K the overall spin-lattice relaxation becomes slower again. For temperatures above the $T_1$ minimum the M(t) curves are nonexponential, see Fig. 3(b), and could be fitted well using the Kohlrausch function

$$M(t) = \exp[-(t/T_1)^{1-\nu}]. \qquad (4)$$

Here the exponent $\nu$ is a measure for the deviation from single exponential behavior. For comparison in Fig. 3(b) we included M(t) as calculated from the $T_1$ data of crystalline aspirin-$d_3$ [11]. This comparison shows that the initial part of the magnetization curves are quite similar for the crystal and the glass but that major deviations exist at long times.

At temperatures near and below the $T_1$ minimum M(t) becomes bimodal and at the lowest temperatures could again be fitted using a Kohlrausch function, see Fig. 3(a). The assignment of the two observed components is not clear at present. In order to obtain a coherent description, the Kohlrausch function was used at all temperatures and thus yielded an estimate of the characteristic time scale and of the overall stretching of M(t). The resulting fitting parameters $T_1$ and $\nu$ are summarized in Fig. 4. Furthermore, the spin-spin relaxation times, $T_2$, are included in this plot. In the range displayed in Fig. 4 the $T_2$ times show little temperature dependence and only for T < 20 K they become significantly shorter than 300 µs.

It is instructive to compare the spin-lattice relaxation times of amorphous aspirin-$d_3$ with those of the crystal [11]. For the latter, the shortest $T_1$ value is $T_{1,min} \approx 3.5$ ms as observed near 52 K at a Larmor frequency of 72.1 MHz. The spin-lattice relaxation times of the crystal were analyzed using the BPP expression

$$\frac{1}{T_1} = 2K \left[ \frac{\tau_C}{1+(\omega_L \tau_C)^2} + \frac{4\tau_C}{1+(2\omega_L \tau_C)^2} \right], \qquad (5)$$

to yield the correlation times $\tau_C$ [11]. The coupling constant is K = 0.702 $\omega_L$ / $T_{1,min} \approx 2\pi \times$ (52 kHz)$^2$. The similar shape of the spectra of the glass and the crystal, cf. Fig. 2, suggests that the coupling constants do not differ much for the two phases. If $T_1$ would follow the BPP expression for the glass as well then its minimum value is expected to be 3 ms × 46.2/72.1 ≈ 2 ms, i.e., about seven times shorter than observed. This discrepancy can be rationalized by assuming the existence of a broad distribution of correlation times characterizing the methyl group motion in the glass. An estimate of the width of this distribution is, however, difficult since it depends sensitively on its shape which is not known. Furthermore, the shape of this distribution is not necessarily of a simple form as a more extensive study on vitreous toluene-$d_3$ has shown [16].

### 3.3 Stimulated-echo experiments

While spin-lattice relaxometry is a suitable tool to study relatively fast motions, the slow dynamics is often accessible by stimulated-echo techniques. Such experiments were sometimes employed to investigate the *classical* slow-down of methyl groups [23], but due to the delocalization of specific CD bonds, in other word the loss of their orientational identity, these experiments are of little use in the quantum mechanical *tunneling* regime. Since this restriction also applies to aspirin-d$_3$, we focus on the slow-down of the molecular reorientation in its glass transformation range. At these relatively high temperatures the motion of the methyl group about its symmetry axis is so fast that the average EFG tensor associated with the C$_3$ axis of the CD$_3$ group is traced. Thus, one can monitor the corresponding two-time autocorrelation function via the damping of the stimulated-echo amplitude

$$F_2(t_p,t_m) \propto \langle \cos[\omega_Q(0)t_p]\cos[\omega_Q(t_m)t_p]\rangle \,, \tag{6}$$

with the quadrupolar frequency $\omega_Q$ given by Eq. (1). In Eq. (6) $t_p$ and $t_m$ denote the evolution and the mixing time, respectively.

Using standard phase cycling [24] we measured the $F_2$ function for several temperatures, see Fig. 5. Since in the experiment the stimulated-echo amplitude is additionally damped by spin-lattice relaxation, the measurements were analyzed using a product of two decay functions, $\propto \exp[-(t_m/\tau_2)^{\beta_2}]\exp[-(t_m/T_1)^{1-\nu}]$. The spin-lattice relaxation time was determined by independent experiments. Since at the temperatures referred to in Fig. 5 spin-lattice relaxation was very close to exponential [7], correlation effects [25] need not to be taken into account in the analysis. The temperature dependence of the stretching exponent $\beta_2$ is shown in Fig. 5. The temperature dependence of the correlation times $\tau_2$ was already discussed in [7] in relation to the time scales determined by other experimental methods.

The evolution time dependence of the correlation times $\tau_2$ is shown in Fig. 6. Since the angular sensitivity of stimulated-echo experiments can be adjusted by $t_p$ these experiments are well suited to map out the jump angles relevant for the molecular reorientation. The relatively strong $t_p$ dependence of $\tau_2$ seen in Fig. 6 demonstrates that small jump angles play an important role for the reorientational dynamics in aspirin-$d_3$ [24]. A more detailed discussion will be given below.

## 4. Discussion

In view of the disorder in glasses it may seem surprising that the low-temperature spectra of amorphous aspirin-$d_3$ look similar to those of the crystal, i.e., that the $CD_3$ tunneling is not sensitive to whether or not its environment is crystalline. However, one should realize that the threefold potential assumed in the calculation of the spectra, cf. the $\cos(3\phi)$ term in Eq. (3), does *not* refer to a possible symmetry in the crystal field. Rather it reflects the permutation symmetry of the deuterons in the methyl group [11]. Consequently, in crystalline aspirin-$CH_2D$ where this symmetry is broken, but the form of the crystal potential is essentially unchanged, isotopic ordering occurs, i.e., the protons and deuterons become localized at low temperatures [17]. The at most ancillary role of the surrounding of a $CD_3$ group justifies the comparison of our spectra with those referring to the crystalline phase of aspirin-$d_3$, cf. Fig. 2.

While the low-temperature spectra of glass and crystal are very similar, the spin-lattice relaxation times revealed major differences. This becomes clear if one realizes that at higher temperatures, i.e., in the classical regime where the correlation rates $\tau_C^{-1}$ are much larger than $\omega_t$, the sensitivity to the local environment of the methyl group is restored. For comparison with the glass data let us note that for the crystal $\tau_C$ was found to follow a thermally activated behavior

$$\tau_C = \tau_0 \exp(\Delta E / T), \qquad (7)$$

with an energy barrier of $\Delta E$ = 510 K (or 44 meV) [11] for T > 40 K. For lower temperatures the effective barrier becomes progressively lower. From the data shown in [11] we estimated a pre-exponential factor of $\tau_0 = 10^{-12}$ s (for T > 40 K). With these parameters the $T_1$ minimum, expected for the slightly lower $\omega_L$ used in our study of the glass, via Eq. (5) is predicted to occur only a few K lower than it does in the crystal. However, the minimum temperatures in the crystal and the glass differ by about 20 K. Therefore, the difference in the Larmor frequencies can not be responsible for this enormous shift. Obviously the energy barrier which dominates the spin relaxation in the glass is significantly lower than it is in the crystal.

In order to estimate the relevant barrier height in the glass let us assume the same pre-exponential factor as in the crystal and the applicability of the relation $\omega_L \tau_C = 0.615$ (valid only for symmetrical distributions of correlation times). Then, we find a barrier of about 250 K, a value considerably lower than in the crystal. This becomes plausible if one regards the broad distribution of barrier heights in the glassy state with significant contributions of (relaxation-relevant) low-energy barriers.

The relatively small barriers relevant in the glass make it obvious that the methyl groups rotate freely near and above the glass transition temperature ($\tau_{methyl} \approx 10^{-11}$ s at 243 K [11]). Hence, it is clear that the stimulated-echo experiments carried out in this range are sensitive to reorientations of the symmetry axis of the methyl group, i.e., to that of the largest principal axis of the average EFG tensor. A rough interpretation of the decay times of the stimulated-echo amplitude presented in Fig. 6 has already been given above. In order to obtain more quantitative information, quasi-analytical calculations [26] as well as random-walk simulations [24] were used to analyze such kind of data. Here we adapt simulations which were previously found to describe the reorientational motion of deeply supercooled glycerol [27] and propylene carbonate [28]. The EFG tensors of these substances are not associated with methyl groups and are characterized by $\eta \approx 0$. Taking into account the difference in the effective coupling constants, i.e., renormalizing the x-axis by a factor of 3 ($\approx \delta_Q/\delta_M$), the previous simulation data are plotted as a solid line in Fig. 6. We point out that nonzero asymmetry parameters have little impact on $\tau_2(t_p)$ except that they lead to a damping of the oscillation seen in the simulated $\tau_2(t_p)$ curve [29]. The simulations plotted in Fig. 6 are based on the assumption of a large (98%) fraction of 2° jumps and a small (2%) fraction of 30° jumps. This bimodal distribution of jump angles provided a simple way to characterize the behavior of several glass formers [27,28] and also reproduces the $\tau_2(t_p)$ data for aspirin-$d_3$ quite well, cf. Fig. 6.

The β-relaxation of glassy aspirin was studied in [6,7] by dielectric spectroscopy. The mean time constants describing the β-process follow Eq. (7) with $\tau_{0,\beta} = 3\times10^{-15}$ s and $\Delta E_\beta = 4900$ K. Hence near $T_g$ the secondary relaxation time, $\tau_\beta(T_g) = 2\times10^{-6}$ s, is about five orders of magnitude longer than the one describing the methyl group rotation. On the other hand, at $T_g$ the α-relaxation time should be considerably longer than $\tau_\beta$. On the basis of his coupling model Ngai has predicted that [30]

$$\tau_\beta = \tau_\alpha^\beta \tau_{cr}^{1-\beta} / \beta. \tag{8}$$

Here β is the relevant stretching of the α-relaxation and $\tau_{cr}$ is a crossover time, which for small-molecule liquids has been reported to be of the order of $2\times10^{-12}$ s [31]. We mention that a relation similar to Eq. (8) (see Eq. (33) in [32]) was derived using a hierarchical model [33]. In this model the relaxation of a number of spins (or molecules) on one level of the hierarchy triggers the relaxation on the next level. Consequently, the corresponding parameters appearing in the expressions derived in the frameworks of the coupling and of the hierarchical approaches have different meanings. In particular, the pendant to $\tau_{cr}$ (called $t_1$ in [32]) refers to the time the system requires to move from one level in the hierarchy to the next. This time was suggested to be independent of a specific hierarchy level, but a numerical value apparently was not specified for $t_1$. In order to test Eq. (8) we extrapolated the experimental stretching exponent for aspirin-$d_3$ to $T_g$ and find $\beta_2 = 0.42$ (cf. Fig. 5). Inserting this value into Eq. (8) gives $\tau_\beta = 2.6\times10^{-6}$ s in excellent agreement with the experimental result for $\tau_\beta(T_g)$ as noted above.

We emphasize that Eq. (8) can thus be taken as a basis for distinguishing Johari-Goldstein β-relaxations from other processes [31]. This property was recently exploited to demonstrate that the only relaxation clearly resolvable below the glass transition temperature of some other pharmaceuticals is *not* of the Johari-Goldstein type [34,35].

## 5. Conclusions

At low temperatures (T < 20 K) the molecular dynamics in glassy aspirin is essentially frozen and rotational tunneling, a quantum mechanical excitation, prevails. Since this process does not depend on the intermolecular environment but only on the symmetry of the methyl group, a large similarity of the spectra of amorphous and crystalline aspirin-$d_3$ was found. At somewhat higher temperatures the classical slow-down of the $CD_3$ group comes into play.

This is clearly seen in the enormous differences showing up when comparing the spin-lattice relaxation times of the glass with those of the crystal. In the glass a wide distribution of the local potentials hindering the methyl motion leads to a broad distribution of spin-lattice relaxation times. In the crystal, on the other hand, the magnetization recovery is exponential. To the extent that the distribution of spin-lattice relaxation times narrows, the typical values of glass and crystal become more similar. In the non-crystalline compound an approximately exponential magnetization recovery is observed near and above the glass transition [7].

For temperatures above $T_g$ the α-relaxation dominates the dynamics. The temperature dependence of the mean times scales governing this and the β-process were reported earlier [5,6,7]. In the present article, we dealt with the reorientational motions associated with the α-process in more detail. In particular, we focused on the width of the distribution of correlation times and on the geometry of the corresponding molecular motion. In both respects the behavior of aspirin was found typical for small molecule glass-forming liquids.

**Acknowledgment**

We wish to thank Kia L. Ngai for numerous stimulating discussions concerning complex systems over the years. This work was supported by the Deutsche Forschungsgemeinschaft within the Graduiertenkolleg 298.

**REFERENCES**

1  J. Bernstein, Polymorphism in Molecular Crystals (Clarendon, Oxford, 2002); see also P. Vishweshwar, J. A. McMahon, M. Oliveira, M. L. Peterson, M. J. Zaworotko, J. Am. Chem. Soc. 127 (2005) 16802.

2  see e.g., L. Yu, Adv. Drug Deliv. Rev. 48 (2001) 27.

3  E. Fukuoka, M. Makita, S. Yamamura, Chem. Pharm. Bull. 37 (1989) 1047.

4  E. Fukuoka, M. Makita, Y. Nakamura, Chem. Pharm. Bull. 39 (1991) 2087.

5  G. P. Johari, D. Pyke, Phys. Chem. Chem. Phys. 2 (2000) 5479.

6  G. P. Johari (private communication).

7  R. Nath, T. El Goresy, B. Geil, H. Zimmermann, R. Böhmer, Phys. Rev. E 74 (2006) 021506.

8  H. Cang, J. Li, H. C. Andersen, M. D. Fayer, J. Chem. Phys. 123 (2005) 064508.

9  A. Nowaczyk, B. Geil, G. Hinze, R. Böhmer, Phys. Rev. E 74 (2006) 041505.

10  M. Kankaanpää, M. Punkkinen, E. E. Ylinen, Mol. Phys. 100 (2002) 287.

11  A. Detken, P. Focke, H. Zimmermann, U. Haeberlen, Z. Olejniczak, Z. T. Lalowicz, Z. Naturforsch. A 50 (1995) 95.

12  Z. Olejniczak, A. Detken, B. Manz, U. Haeberlen, J. Magn. Reson. A 118 (1996) 55.

13  S. Szymański, Z. Olejniczak, A. Detken, U. Haeberlen, J. Magn. Reson. 148 (2001) 277.

14  For a review see A. J. Horsewill, Prog. NMR Spectrosc. 35 (1999) 359.

15  K. Börner, G. Diezemann, E. Rössler, H.-M. Vieth, Chem. Phys. Lett. 181 (1991) 563.

16  F. Qi, G. Hinze, R. Böhmer, H. Sillescu, H. Zimmermann, Chem. Phys. Lett. 328 (2000) 257.

17  P. Schiebel, R. J. Papoular, W. Paulus, H. Zimmermann, A. Detken, U. Haeberlen, W. Prandl, Phys. Rev. Lett. 83 (1999) 975.

18  S. J. Kitchin, T. K. Halstead, Appl. Magn. Reson. 17 (1999) 283.

19  E. Rössler, M. Taupitz, K. Börner, M. Schulz, H.-M. Vieth, J. Chem. Phys. 92 (1990) 5847.

20 V. Macho, L. Brombacher, H. W. Spiess, Appl. Magn. Reson. 20 (2001) 405.

21 Z. T. Lalowicz, U. Werner, W. Müller-Warmuth, Z. Naturforsch. A 43 (1988) 219.

22 G. Hinze, Ph.D. thesis, Universität Mainz, 1993, citing: G. Diezemann (private communication).

23 F. Qi, R. Böhmer, H. Sillescu, Phys. Chem. Chem. Phys. 3 (2001) 4022.

24 R. Böhmer, G. Diezemann, G. Hinze, E. Rössler, Prog. NMR Spectrosc. 39 (2001) 191.

25 R. Böhmer, G. Diezemann, B. Geil, G. Hinze, A. Nowaczyk, M. Winterlich, Phys. Rev. Lett. 97 (2006) 135701.

26 B. Geil, F. Fujara, H. Sillescu, J. Magn. Reson. 130 (1998) 18.

27 R. Böhmer, G. Hinze, J. Chem. Phys. 109 (1998) 241.

28 F. Qi, K. U. Schug, A. Döß, S. Dupont, R. Böhmer, H. Sillescu, H. Kolshorn, H. Zimmermann, J. Chem. Phys. 112 (2000) 9455.

29 U. Tracht, A. Heuer, H. W. Spiess, J. Chem. Phys. 111 (1999) 3720.

30 K. L. Ngai, Comments Solid State Phys. 9 (1979) 127; *ibid*. 9 (1979) 149.

31 see e.g., K. L. Ngai, M. Paluch, J. Chem. Phys. 120 (2004) 857.

32 J. Y. Cavaille, J. Perez, G. P. Johari, Phys. Rev. B 39 (1989) 2411.

33 R. J. Palmer, D. L. Stein, E. Abrahams, P. W. Anderson, Phys. Rev. Lett. 53 (1984) 958.

34 T. El Goresy, R. Böhmer, J. Phys. - Condensed Matter 19 (2007) in press.

35 T. El Goresy, R. Böhmer, J. Non-Cryst. Solids 352 (2006) 4459.

**FIGURE CAPTIONS**

**Fig. 1**

Solid-echo spectra of aspirin-$d_3$ recorded using an echo delay of 50 µs. The dashed line serves to emphasize the feature at –63.5 kHz. The dotted line is calculated [20] using $\delta_M = 2\pi \times 43.5$ kHz, $\eta_M = 0.1$, and a Lorentzian broadening of 3 kHz.

**Fig. 2**

The experimental spectrum of aspirin-$d_3$ obtained at 17.2 K (noisy line) is compared to various theoretical spectra. In the upper part of the figure the simulations performed for a polycrystal ($e^2qQ/\hbar$ = 170 kHz and $\omega_t/2\pi$ = 2.59 MHz) are adapted from [12]. The uppermost trace corresponds to the calculated FID spectrum ($t_{SE}$ = 0), then calculated solid-echo spectra with pulse delays of $t_{SE}$ = 26, 42, and 58 µs follow. The dashed line represents a superposition of a Pake pattern ($\delta_Q = 2\pi \times 131$ kHz, $\eta = 0$) with the theoretical spectrum shown in Fig. 1.

**Fig. 3**

Normalized magnetization recoveries M(t) measured for glassy aspirin-$d_3$. Frame (a) shows that at low temperatures M(t) can be fitted by a superposition of two Kohlrausch functions. The dashed line is a fit using one such function, cf. Eq. (4). Frame (b) compares M(t) for the glass (symbols and solid lines) with the exponential magnetization curves (dotted lines) calculated from the $T_1$ data [11] of the crystal at the same temperatures. While the initial decays for glass and crystal are rather similar, the deviations become very pronounced for longer times, particularly at low temperatures.

**Fig. 4**

(a) Spin-relaxation times, $T_1$, for crystalline aspirin-$d_3$ (open symbols, [11]) in comparison with those of the glass (solid symbols). (b) Stretching exponents ν obtained for the glass from fits using Eq. (4). Note that for T ≤ 50 K systematic deviations from the Kohlrausch function exist, cf. Fig. 3, and that sometimes the exponent ν was fixed (diamonds).

**Fig. 5**

The stimulated-echo amplitude for supercooled aspirin-$d_3$ recorded as a function of the mixing time at several temperatures. The symbols represent measurements that were obtained at an evolution time $t_p$ of 25 µs. The lines are fits using a Kohlrausch function as described in the text. The inset shows the temperature dependence of the Kohlrausch exponent $\beta_2$. The straight line is used to extrapolate $\beta_2$ to $T_g$ which is marked by an arrow.

**Fig. 6**

Stimulated-echo decay times $\tau_2$ for supercooled aspirin-$d_3$ are shown as a function of the evolution time $t_p$. The experimental data obtained for two temperatures are represented by the symbols. The solid line is the result of a random-walk simulation as described in the text.

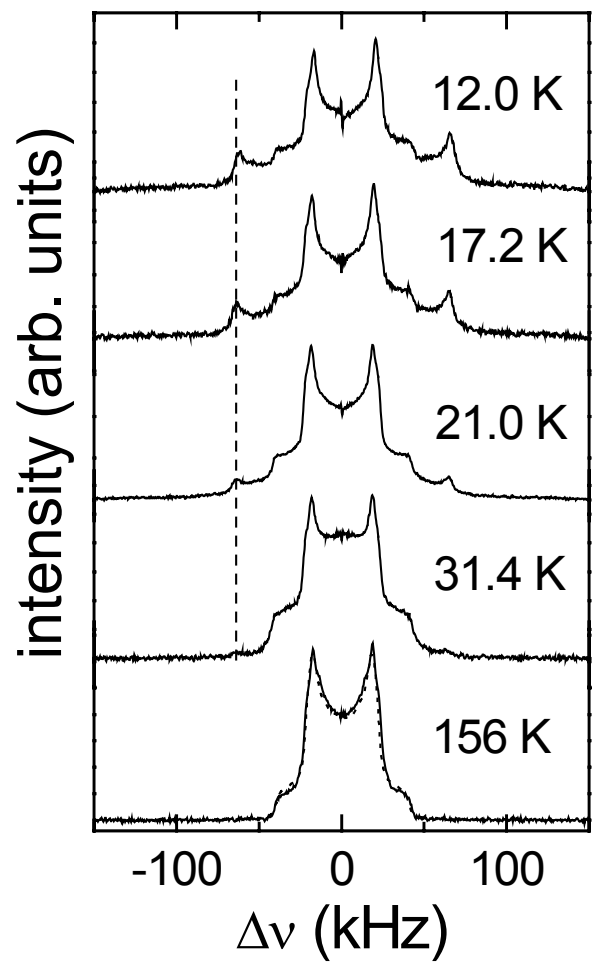

Fig. 1

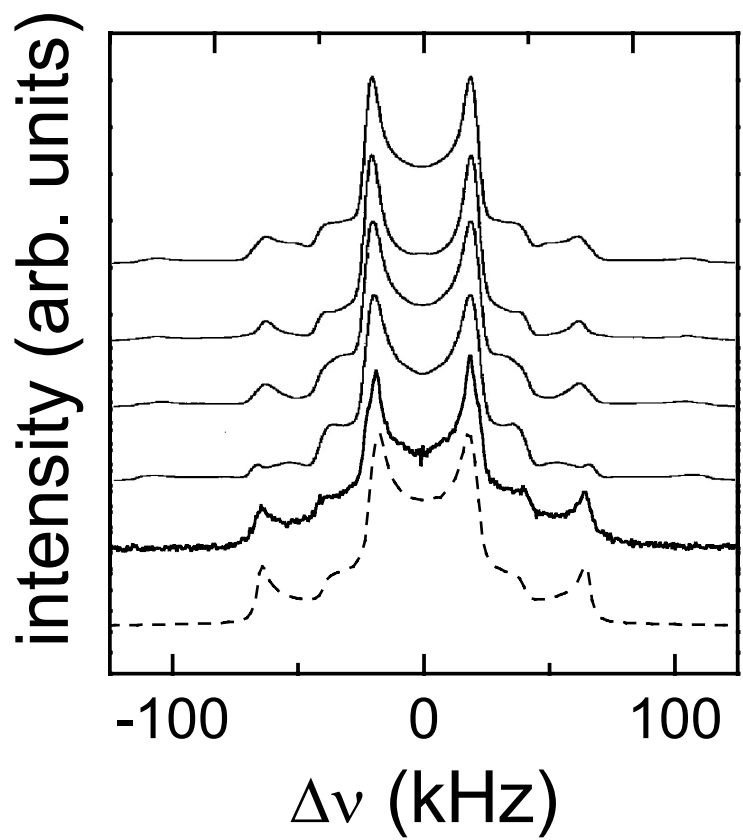

Fig. 2

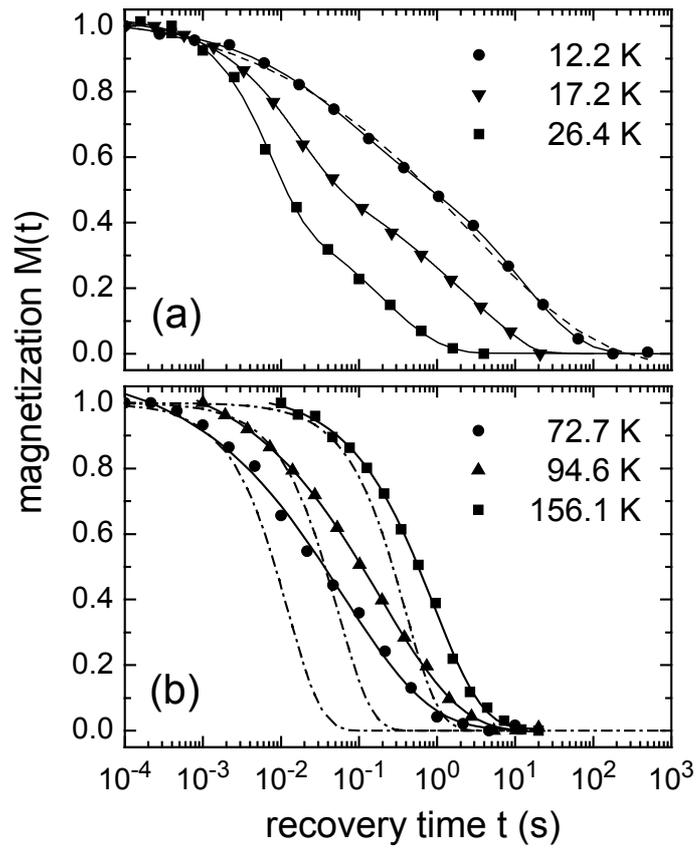

Fig. 3

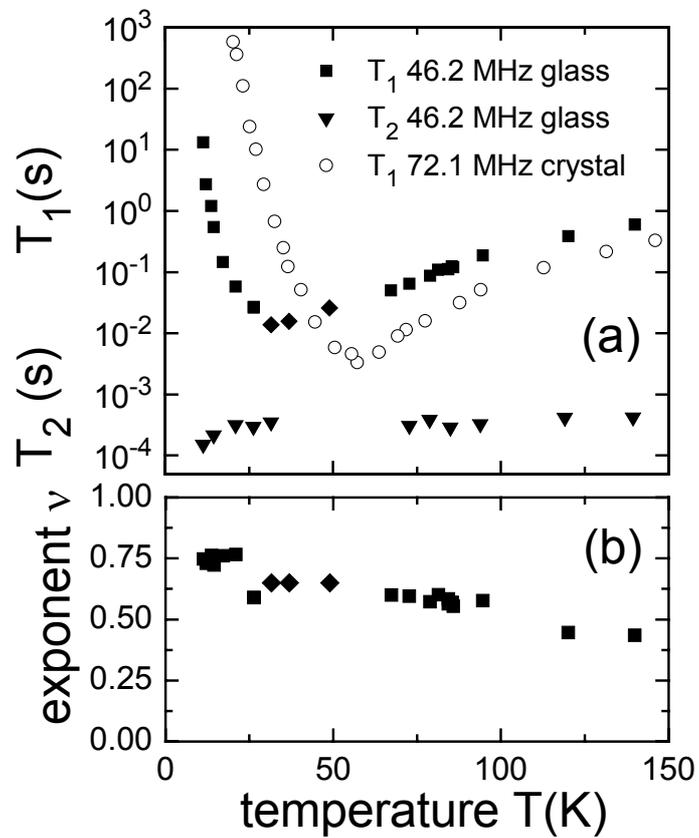

Fig. 4

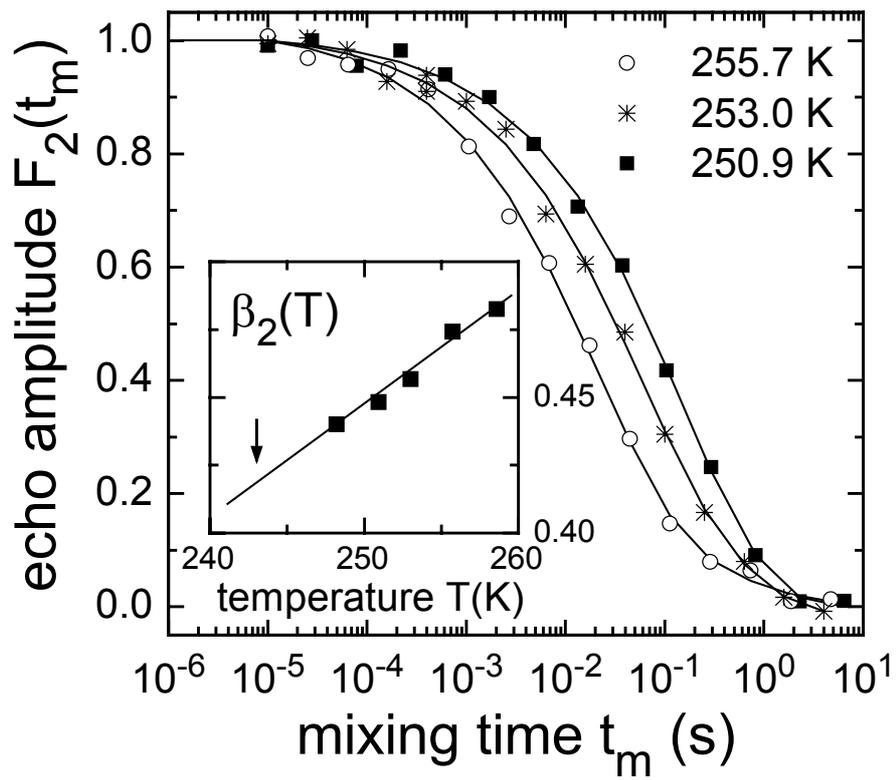

Fig. 5

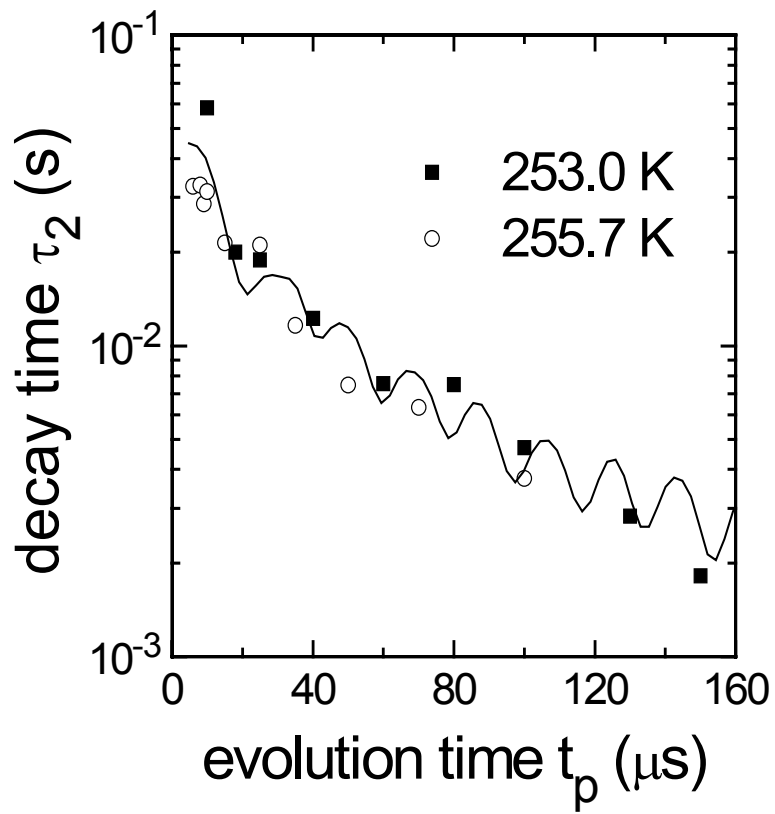

Fig. 6